\documentclass{aip-cp}

\usepackage[numbers]{natbib}
\usepackage{rotating}
\usepackage{graphicx}

\begin{document}


\title{Longitudinal vector form factors in weak decays of nuclei}

\author[aff3,aff4]{F. \v{S}imkovic}
\author[aff5]{S. Kovalenko}
\author[aff3,aff1]{M. I. Krivoruchenko}
\affil[aff3]{Bogoliubov Laboratory of Theoretical Physics, JINR, 141980 Dubna, Russia}
\affil[aff4]{Department of Nuclear Physics and Biophysics, Comenius University, Mlynsk\'a dolina F1 \\
SK--842 48 Bratislava, Slovakia}
\affil[aff5]{Universidad T\'{e}cnica Federico Santa Mariya,
Centro-Cientifico-Tecnol\'{o}gico de Valparaiso,
Casilla 110-V, Valparaiso, Chile}
\affil[aff1]{Institute for Theoretical and Experimental Physics, B. Cheremushkinskaya 25, 117218 Moscow, Russia}

\maketitle

\begin{abstract}
The longitudinal form factors of the weak vector current of particles with spin $ J = 1/2 $ and isospin $ I = 1/2 $ are determined by the mass difference and the charge radii of members of the isotopic doublets. The most promising reactions to measure these form factors are the reactions with large momentum transfers involving the spin-1/2 isotopic doublets with a maximum mass splitting. Numerical estimates of longitudinal form factors are given for nucleons and eight nuclear spin-1/2 isotopic doublets.
\end{abstract}

\vspace{5pt}
$\;$

The weak vector current (VC) has the transformation properties of the isotopic triplet. Its 3rd component is associated with the electromagnetic current. The weak VC can be recovered from the electromagnetic current by symmetry considerations. Gerstein and Zeldovich \cite{Gerstein1955} and Feynman and Gell-Mann \cite {Feynman1958} used isotopic rotation to obtain tensor component of the weak VC in $\beta$ decay amplitudes. This component is determined by the difference between the anomalous magnetic moments of members of the isotopic doublet.

On the mass shell, when the masses of the initial and final states are the same, the vector current of spin $ J = 1/2 $ particles is parameterized by two form factors. Off the mass shell the current has a more complex structure. In general, the renormalized vector vertex is described by 12 form factors
\cite{Binc1960,Tiem1990}:
\begin{equation} \label{GEQ103}
\Gamma_{\mu } (p^{\prime},p) = \sum _{\kappa^{\prime} \kappa } \Lambda _{\kappa^{\prime}} (p^{\prime})(\gamma _{\mu } {\rm {\mathcal F}}_{1}^{\kappa^{\prime} \kappa } +i\sigma _{\mu \nu } q_{\nu } {\rm {\mathcal F}}_{2}^{\kappa^{\prime}\kappa } +q_{\mu } (p^{\prime 2} -p^{2} ){\rm {\mathcal F}}_{3}^{\kappa^{\prime}\kappa } )\Lambda _{\kappa } (p),
\end{equation}
where $\kappa^{\prime}$, $\kappa$ take values $\pm 1$, $q=p^{\prime}-p$ is the momentum transfer,
\[
\Lambda _{\kappa } (p)=\frac{\kappa \hat{p}+M}{2M}
\]
are projection operators to states with the virtual mass $M=\sqrt{p^{2}} $, $\hat{p} \Lambda_{\kappa } (p)=\kappa M \Lambda _{\kappa } (p)$, and ${\rm {\mathcal F}}_{\alpha }^{\kappa^{\prime} \kappa } $ are functions of three independent variables $p^{\prime 2}$, $p^{2}$ and $q^{2} $. The negative $C$-parity of the vertex
\begin{equation} \label{GEQ102bis}
C^{T} \Gamma _{\mu } (p^{\prime},p)^{T} C=-\Gamma _{\mu } (p^{\prime},p),
\end{equation}
where $C=i\gamma ^{2} \gamma _{0} $, by virtue of $\gamma _{\mu } \mathop{\to }\limits^{C} -\gamma _{\mu } $, $\sigma _{\mu \nu } \mathop{\to }\limits^{C} -\sigma _{\mu \nu } $, $(p^{\prime},p) \mathop{\to }\limits^{C} (-p,-p^{\prime})$, implies
\begin{equation} \label{GEQ102bisbis}
{\rm {\mathcal F}}_{\alpha }^{\kappa^{\prime}\kappa } (p^{\prime 2} ,p^{2} ,q^{2} )={\rm {\mathcal F}}_{\alpha }^{\kappa \kappa^{\prime}} (p^{2} ,p^{\prime 2} ,q^{2} ).
\end{equation}
${\rm {\mathcal F}}_{\alpha }^{\kappa \kappa } $ are symmetric functions of the first two variables.
${\rm {\mathcal F}}_{3}^{++} (p^{\prime 2} ,p^{2} ,q^{2} )$ is in general different from zero, but in the limit of exact isotopic symmetry its contribution to physical amplitudes can vanish because of the factor $ p^{\prime 2} - p^{2}$.
Similar properties hold for the longitudinal vector form factor of the pion $\beta$ decay \cite{MIK14}.

\vspace{15pt}

We start from the case of equal fermion masses $m_{f} =m_{i} \equiv m$. The Ward-Takahashi identity (WTI) imposes constraint on the vertex function
\begin{equation} \label{WardId}
q_{\mu } \Gamma _{\mu } (p^{\prime},p)=S^{-1} (p^{\prime})-S^{-1} (p),
\end{equation}
where $S(p)$ is the renormalized propagator. For the vertex (\ref{GEQ103}) WTI gives
\begin{eqnarray}
&&\Lambda _{\kappa^{\prime}} (p^{\prime})q^{2} (p^{\prime 2} -p^{2} ){\rm {\mathcal F}}_{3}^{\kappa^{\prime}\kappa } (p^{\prime 2} ,p^{2} ,q^{2} )
\Lambda _{\kappa }(p) \nonumber \\
&& \;\;\;\;\;\;\;\;\;\;\;\;\;\;\;\;\;\;\;\;\; =\Lambda _{\kappa^{\prime}} (p^{\prime})\left(-\hat{q}{\rm {\mathcal F}}_{1}^{\kappa^{\prime}\kappa } (p^{\prime 2} ,p^{2} ,q^{2} )+S^{-1} (p^{\prime})-S^{-1} (p)\right) \Lambda _{\kappa } (p).
\label{GEQ104}
\end{eqnarray}
The inverse renormalized propagator can be represented in the form
\begin{equation} \label{GEQ105}
S^{-1}(p) = \hat{p} - m - \Sigma.
\end{equation}
The mass operator, $\Sigma$, expands over the gamma-matrices:
$
\Sigma =(\hat{p}-m)\Sigma _{1}+\Sigma _{2},
$
where $\Sigma _{1}$ and $\Sigma _{2}$ are scalar functions of $p^2$ and $m$.
Given that $m$ is the physical mass and the residue in $p^2$ at the pole of $S(p)$ equals $\hat{p} + m$, we obtain on the mass shell
\begin{equation} \label{reno}
\Sigma _{1} = \Sigma _{2} = \frac{\partial}{\partial p^{2}} \Sigma _{2} = 0.
\end{equation}
The last equation is satisfied for $ \Sigma_{2} = (p^{2} -m^{2})^{2} \Sigma_{3} $, where $ \Sigma_{3} $ is arbitrary scalar function of $ p^2 $ and
$ m $. From this representation, it follows
\begin{equation}
\frac{\partial}{\partial m} \Sigma _{2} = 0
\end{equation}
and thus, for $p^{2}=m^{2}$,
\begin{eqnarray}
\Sigma &=& 0, \label{Sigma0} \\
\Lambda_{+}(p) \frac{\partial}{\partial m} \Sigma  &=& 0, \label{dSigma0} \\
\Lambda_{+}(p) \frac{\partial}{\partial m} S^{-1}(p)  &=& - \Lambda_{+}(p). \label{dS}
\end{eqnarray}

Of interest is the positive energy component of the current. Substituting the expression (\ref{GEQ105}) in (\ref{GEQ104}),
we find for $ \kappa = \kappa^{\prime} = + 1 $
\begin{eqnarray}
q^{2} \left(M'^{2} -M^{2} \right){\rm {\mathcal F}}_{3}^{++} (M'^{2} ,M^{2} ,q^{2} )&=&
-( M'- M ){\rm {\mathcal F}}_{1}^{++} (M'^{2} ,M^{2} ,q^{2} )
\nonumber \\
&&+ M^{\prime} - M \nonumber \\
&&- \left( (M^{\prime} - m)\Sigma_{1}(M^{\prime 2},m) + \Sigma_{2}(M^{\prime 2},m) \right) \nonumber \\
&&+ \left( (M - m)\Sigma_{1}(M^{2},m) + \Sigma_{2}(M^{2},m) \right). \label{GEQ109}
\end{eqnarray}
This equation is fulfilled for $M'=M$. We take the derivative on $M'$ and set $M'=M=m$. Using (\ref{reno}) we obtain
\begin{equation} \label{GEQ106}
{\rm {\mathcal F}}_{3}^{++} (m^{2} ,m^{2} ,q^{2} )
= - \frac{{\rm {\mathcal F}}_{1}^{++} (m^{2} ,m^{2} ,q^{2} )-1}{2mq^{2} } .
\end{equation}
In the neighborhood of $q^{2} = 0 $, ${\rm {\mathcal F}}_{1}^{++} (m^{2} ,m^{2} ,q^{2} )$ can be expanded to give $1 + q^2\left\langle r^{2} \right\rangle /6 + \ldots$,
where $\left\langle r^{2} \right\rangle $ is the Lorentz-vector charge radius of the particle (in units of the proton electric charge). For low $q^{2} $
\begin{equation} \label{GEQ107}
{\rm {\mathcal F}}_{3}^{++} (m^{2} ,m^{2} ,0)=-\frac{1}{12m} \left\langle r^{2} \right\rangle .
\end{equation}

\vspace{15pt}

In the isotopic doublet $ ((A, Z + 1), (A, Z)) $, the rising component of the weak vector vertex $2 \tau ^{+} \Gamma _{v \mu } (p^{\prime},p)$
has the Lorentz structure of the expression (\ref{GEQ103}). To get  $\Gamma _{v\mu } (p^{\prime},p)$, one has to make the substitution
${\rm {\mathcal F}}_{\alpha }^{\kappa^{\prime}\kappa } $ $\to {\rm {\mathcal F}}_{\alpha v}^{\kappa^{\prime}\kappa } $
$=({\rm {\mathcal F}}_{\alpha {I_{3} = +1/2} }^{\kappa^{\prime}\kappa } -{\rm {\mathcal F}}_{\alpha {I_{3} = -1/2}}^{\kappa^{\prime}\kappa } )/2.$ The positive-energy Lorentz-vector isovector form factors are normalized by the conditions:
\begin{eqnarray}
2 {\rm {\mathcal F}}_{1v}^{++} (m^{2} ,m^{2} ,0)&=&1, \label{norm1} \\
2 {\rm {\mathcal F}}_{2v}^{++} (m^{2} ,m^{2} ,0) &=& \Delta \mu_{I_{3} = +1/2} - \Delta \mu_{I_{3} = -1/2}, \label{norm2} \\
-24m {\rm {\mathcal F}}_{3v}^{++} (m^{2} ,m^{2} ,0)&=&\left\langle r^{2} \right\rangle _{I_{3} = +1/2} -\left\langle r^{2} \right\rangle _{I_{3} = -1/2}, \label{norm3}
\end{eqnarray}
where $\Delta \mu_{I_{3}}$ and $\left\langle r^{2} \right\rangle _{I_{3}}$ are anomalous magnetic moments and Lorentz-vector charge radii of the components of the isotopic doublet. The 1-st isoscalar form factor is normalized to  ${\rm {\mathcal F}}_{1s}^{++} (m^{2} ,m^{2} ,0)= Z + 1/2$,
where $Z$ is the charge of the $I_{3} = - 1/2$ component. In the isovector channel, the relation (\ref{GEQ106}) reads
\begin{equation} \label{GEQ106bis}
2{\rm {\mathcal F}}_{3 v}^{++} (m^{2} ,m^{2} ,q^{2} )
= - \frac{2{\rm {\mathcal F}}_{1v}^{++} (m^{2} ,m^{2} ,q^{2} )-1}{2mq^{2} } .
\end{equation}

The coupling to the longitudinal component of the weak VC in the $\beta$-decay amplitude to the first order in the mass difference between the initial and final particles can be represented as
\begin{eqnarray}
\frac{g_{v} }{m_{\mu } } &\equiv& (m_{I_{3} = +1/2}^{2} -m_{I_{3} = -1/2}^{2} )2{\rm {\mathcal F}}_{3v}^{++} (m^{2} ,m^{2} ,0) \nonumber \\
&=&-\frac{m_{I_{3} = +1/2}^{} -m_{I_{3} = -1/2}^{} }{6} \left(\left\langle r^{2} \right\rangle _{I_{3} = +1/2} -\left\langle r^{2} \right\rangle _{I_{3} = -1/2} \right), \label{GEQ108}
\end{eqnarray}
where $m_{\mu } $ is the muon mass, taken for the normalization.

Isotopic rotation alone is, however, not enough to obtain a full weak-interaction vertex. The violation of isotopic symmetry generates a Lorentz-scalar isovector contribution that is unrelated to isotopic rotation.

\vspace{15pt}

WTI (\ref{WardId}) holds when the symmetry is exact. The estimate (\ref{GEQ108}) neglects corrections related to the isotopic symmetry breaking. We restrict ourselves to first order in the mass difference
$ \Delta m_{fi} = m_{f} - m_{i} $ and consider the isovector vertex $ \Gamma_{fi}^{\mu} (p^{\prime}, p) $, associated with the rising component of the weak VC. In the vertex the fermion masses are changed. As shown in Ref.~\cite{PEPAN2015}, for a sufficiently broad class of meson-nucleon effective theories with broken isotopic symmetry the generalized WTI takes place:
\begin{equation} \label{GEQ110}
q_{\mu } \Gamma _{fi}^{\mu } (p^{\prime},p)=S_{f}^{-1} (p^{\prime})-S_{i}^{-1} (p)+\delta m_{fi} \Theta _{fi} (p^{\prime},p),
\end{equation}
where $S_{i}(p)$ and $S_{f}(p)$ are the renormalized nucleon propagators with the masses $m_{i} $ and $m_{f} $ in the initial and final states,
$\Theta_{fi} (p^{\prime},p)$ is the rising component of the Lorentz-scalar isovector vertex. Eq.~(\ref{GEQ110}) holds to the first order in the weak interaction. We expand the scalar vertex over the projection operators
\begin{equation} \label{GEQ111}
\Theta _{fi} (p^{\prime},p)=2 \sum _{\kappa^{\prime}\kappa } \Lambda _{\kappa^{\prime}} (p^{\prime}){\rm {\mathcal F}}_{7v}^{\kappa^{\prime}\kappa } \Lambda _{\kappa } (p),
\end{equation}
and represent the vector vertex in the form
\begin{equation} \label{GEQ112bis}
\Gamma _{fi}^{\mu } (p^{\prime},p)= 2 \left(\Gamma _{v}^{\mu } (p^{\prime},p)+\Gamma _{1}^{\mu } (p^{\prime},p)\right).
\end{equation}
To the first order in the mass difference $\delta m_{fi} $ the most general representation for the second term is as follows
\begin{equation} \label{GEQ112bisbis}
\Gamma _{1}^{\mu } (p^{\prime},p)=\delta m_{fi} \sum _{\kappa^{\prime}\kappa } \Lambda _{\kappa^{\prime}} (p^{\prime})\left(\gamma _{\mu } (p^{\prime 2} -p^{2} ){\rm {\mathcal F}}_{4v}^{\kappa^{\prime}\kappa } +i\sigma _{\mu \nu } q_{\nu } (p^{\prime 2} -p^{2} ){\rm {\mathcal F}}_{5v}^{\kappa^{\prime}\kappa } +q_{\mu } {\rm {\mathcal F}}_{6v}^{\kappa^{\prime}\kappa } \right)\Lambda _{\kappa } (p).
\end{equation}
By interchanging the initial and final states the sign of $\delta m_{fi} $ gets opposite, so under the conditions
$
{\rm {\mathcal F}}_{\alpha v}^{\kappa^{\prime}\kappa } (p^{\prime 2} ,p^{2} ,q^{2} )={\rm {\mathcal F}}_{\alpha v}^{\kappa \kappa^{\prime}} (p^{2} ,p^{\prime 2} ,q^{2} )
$
all terms in Eq.~(\ref{GEQ112bisbis}) have positive $G$-parity.
On the mass shell, the contributions of ${\rm {\mathcal F}}_{4v}^{\kappa \kappa } $ and ${\rm {\mathcal F}}_{5v}^{\kappa \kappa } $
have additional smallness in $\delta m_{fi} $ and, therefore, there two form factors can be neglected.

To the first order of the expansion, Eq.~(\ref{GEQ110}) gives
\[
2 \Lambda _{\kappa^{\prime}} (p^{\prime})\left(q^{2} {\rm {\mathcal F}}_{6v}^{\kappa^{\prime}\kappa } (p^{\prime 2} ,p^{2} ,q^{2} )-{\rm {\mathcal F}}_{7v}^{\kappa^{\prime}\kappa } (p^{\prime 2} ,p^{2} ,q^{2} )\right)\Lambda _{\kappa } (p)=
\Lambda _{\kappa^{\prime}} (p^{\prime}) \frac{1}{2} \frac{\partial}{\partial m} \left(S^{-1} (p^{\prime})+S^{-1} (p)\right) \Lambda _{\kappa} (p).
\]
We consider the positive frequency component on the mass shell $M' = m_{f}$, $M = m_{i}$. Taking into account (\ref{dS}), one finds
\begin{equation} \label{GEQ112}
2{\rm {\mathcal F}}_{6v}^{++} (m^{2}_{f} ,m^{2}_{i} ,q^{2} ) =
\frac{2{\rm {\mathcal F}}_{7v}^{++} (m^{2}_{f} ,m^{2}_{i} ,q^{2} )-1}{q^{2} }.
\end{equation}
The normalization ${\rm {\mathcal F}}_{7v}^{++}(q^2 = 0) = 1$ implies that $ {\rm {\mathcal F}}_{6v}^{++} $ is determined, at small momentum transfers, by the isovector Lorentz-scalar charge radius.
\vspace{15pt}

In the non-relativistic theory Lorentz nature of the charge radii of bound states is not important, because the vector charge radius coincides with the scalar charge radius. In the longitudinal component of the weak VC, contributions (\ref{GEQ106bis}) and (\ref{GEQ112}) for small $ q^{2} $ cancel each other. The effect of the Fermi motion of nucleons in the nuclei is of the order of $ (v/c)^{2}$, where $v$ is velocity of nucleons, and therefore small. The leading contribution comes from the charge radii of nucleons, since nucleons are composed of relativistic quarks.

In the MIT bag model \cite{ChodosA1974} the vector and scalar densities of the quarks are normalized by the conditions
\begin{eqnarray}
1 &=&\int d\mathbf{x}\psi _{i}^{\dagger }\psi _{i}, \label{v1} \\
\langle 1_{s}\rangle  &=&\int d\mathbf{x}\bar{\psi}_{i}\psi _{i}=0.479. \label{v2}
\end{eqnarray}
The value $\langle 1_{s}\rangle$ is the nucleon scalar charge. The corresponding charge radii equal
\begin{eqnarray}
\langle r^{2}\rangle  &=& \sum_{i} e_{i} \int d\mathbf{x} \psi_{i}^{\dagger }\mathbf{x}^{2}\psi
_{i}= \sum_{i} e_{i}  0.531R^{2}, \\
\langle r_{s}^{2}\rangle  &=& \sum_{i} e_{i} \int d\mathbf{x}\bar{\psi}_{i}\mathbf{x}%
^{2}\psi _{i} = \sum_{i} e_{i} 0.182R^{2},
\end{eqnarray}
where $R$ is the bag radius, $e_{i} $ are the quark charges in the unit of the proton charge.
The values $\left\langle r^{2} \right\rangle$ è $\left\langle r^{2}_{s} \right\rangle$
enter the expansion of the form factors at small $q^{2}$. Given the quark wave functions, one may find vector and scalar form factors
\begin{eqnarray}
\sum_{i} \tau_{i}^{+} 2{\rm {\mathcal F}}_{1v}^{++} &=& \sum_{i}\int d\mathbf{x}\psi_{i}^{\dagger} \tau_{i}^{+} \psi_{i}e^{i\mathbf{qx}}, \label{ffv} \\
\langle 1_{s} \rangle \sum_{i} \tau_{i}^{+} 2{\rm {\mathcal F}}_{7v}^{++} &=& \sum_{i}\int d\mathbf{x}\bar{\psi}_{i}\tau_{i}^{+} \psi_{i}e^{i\mathbf{qx}}. \label{ffs}
\end{eqnarray}
The normalization of the vector form factor (\ref{norm1}) is consistent with the normalization of the quark wave functions (\ref{v1}), so the equation (\ref{ffv}) for $ q^2 = 0 $ is an identity. The scalar form factor $ 2 {\rm {\mathcal F}}_{7v}^{++} $ also is normalized to unity, however, the right-hand side of Eq.~(\ref{ffs}) for
$ q^2 = 0 $ equals $ \left \langle 1_{s} \right \rangle \sum_ {i} \tau_{i}^{+} $, so the factor $ \langle 1_{s} \rangle $ is added to the left-hand side. The first expansion coefficient of $ 2 {\rm {\mathcal F}}_{7v}^{++} $ in powers of $ q^2 $ is defined by the scalar charge radius in the units of $ \langle 1_{s} \rangle $
\begin{equation}
\langle r_{s}^{2} \rangle / \langle 1_{s} \rangle = \sum_{i} e_{i} 0.380 R^{2}.
\end{equation}

In the case of nucleons, we get
\begin{equation} \label{GEQ114}
\frac{g_{v+s}}{m_{\mu }} = - \frac{m_{p}^{} -m_{n}^{} }{6} \left[\left(\left\langle r^{2} \right\rangle _{p}^{} -\left\langle r^{2} \right\rangle _{n}^{} \right)-\left(\left\langle r_{s}^{2} \right\rangle _{p} -\left\langle r_{s}^{2} \right\rangle _{n} \right)/ \langle 1_{s} \rangle  \right].
\end{equation}

\begin{table}[t]
\caption{
The longitudinal vector form factors of the weak VC in the $\beta$ decay of $ I = J = 1/2 $ nuclear isotopic doublets.
Q-value and the mass difference $ \Delta m = m_ {I_ {3} = -1/2} - m_ {I_ {3} = + 1/2} $ are shown in the columns 4 and 5.
$ g_ {v + s} ^ {{\rm th}} $ are the predicted values of the form factors at zero momentum transfer.
The experimental value of $ g_{v + s}^{{\rm expt}} $ for the $ (^{3} {\rm He}, ^{3} {\rm H)} $ is given in Ref.~\cite {Gazit2008} .
$ ^{33} {\rm S}^{{\rm *}} $ and $ ^{69} {\rm Ga}^{{\rm *}} $ are excited nuclear states; it is assumed that they are components isotopic doublets.
Except for the last two lines, all the pairs are mirror nuclei.
}
\label{tab:1}
\centering
\vspace{2pt}
\begin{tabular}{|l|l|c|r|r|r|c|}
\hline
$I_{3} =-1/2$  & $I_{3} =+1/2$ & $J^{\pi } $ & $Q$~[keV]~~~~ & $\Delta m $~[keV] & $g_{v+s}^{{\rm th}}~~~~ $ & $g_{v+s}^{{\rm expt}} $ \\
\hline
 ~~$_{0}^{1}   {\rm n}_{1} $   &  ~~$_{1}^{1} {\rm p}_{0}    $  & $1/2^+$  & 782.354 ($\beta^-$)  &  $1~293.353$ & $ 1.1 \times 10^{-4}$  &  \\ \hline
 ~~$_{1}^{3}   {\rm H}_{2} $   &  ~~$_{2}^{3} {\rm He}_{1}   $  & $1/2^+$ & 18.591 ($\beta^-$)    &  $  529.590$ & $ 4.4 \times 10^{-5}$  & $-0.005 \pm 0.040$  \\ \hline
 ~~$_{6}^{13}  {\rm C}_{7} $   &  ~~$_{7}^{13} {\rm N}_{6}   $  & $1/2^-$ &  2 220.4 (EC)         & $-2~220.4$   & $-1.8 \times 10^{-4}$  &  \\ \hline
 ~~$_{7}^{15}  {\rm N}_{8} $   &  ~~$_{8}^{15} {\rm O}_{7}   $  & $1/2^-$ &  2 753.9 (EC)         & $-2~753.9$   & $-2.3 \times 10^{-4}$  &  \\ \hline
 ~~$_{9}^{19}  {\rm F}_{10} $  &  ~~$_{10}^{19} {\rm Ne}_{9} $  & $1/2^+$ &  3.238.4 (EC)         & $-3~238.4$   & $-2.7 \times 10^{-4}$  &  \\ \hline
 ~~$_{14}^{29} {\rm Si}_{15}$  &  ~~$_{15}^{29} {\rm P}_{14} $  & $1/2^+$ &  4943.1 (EC)          & $-4~943.1$   & $-4.1 \times 10^{-4}$  &  \\ \hline
 ~~$_{15}^{31} {\rm P}_{16} $  &  ~~$_{16}^{31} {\rm S}_{15} $  & $1/2^+$ &  5396.1 (EC)          & $-5~396.1$   & $-4.5 \times 10^{-4}$  &  \\ \hline
 ~~$_{15}^{33} {\rm P}_{18} $  &  ~~$_{16}^{33} {\rm S}_{17}^{*}  $  & $1/2^+$ &  284.5 ($\beta^-$) & $-81.5$    & $-6.7 \times 10^{-6}$  &  \\ \hline
 ~~$_{30}^{69} {\rm Zn}_{39}$  &  ~~$_{31}^{69} {\rm Ga}_{38}^{*} $  & $1/2^-$ &  906.0 ($\beta^-$)  & $1~099.0$  & $ 9.0 \times 10^{-5}$  &  \\
\hline
\end{tabular}%
\end{table}

The maximum effect of the longitudinal component of the weak VC is expected in the processes with relatively large momentum transfer, e.g., in the muon capture. Let us consider the isotopic doublet $ ((A,Z + 1), (A,Z))$. The vector and scalar charge radii of nucleus $ (A,Z) $ can be written as
\begin{eqnarray}
\left\langle r^{2}\right\rangle _{\mathrm{(A,Z)}} &=&Z\left\langle
r^{2}\right\rangle _{p}+(A-Z)\left\langle r^{2}\right\rangle
_{n}+\sum_{i=1}^{Z}\langle r_{i}^{2}\rangle , \label{rv} \\
\left\langle r_{s}^{2}\right\rangle _{\mathrm{(A,Z)}} &=& Z\left\langle
r_{s}^{2}\right\rangle _{p}+(A-Z)\left\langle r_{s}^{2}\right\rangle
_{n}+ \left\langle 1_{s}\right\rangle \sum_{i=1}^{Z}\langle r_{i}^{2}\rangle .
\label{rs}
\end{eqnarray}
In both cases, the first two terms describe the contribution of the nucleon charge radii, the last term takes into account the structure of the nucleus, $ r_{i} $ is the proton distance from the center of gravity of the nucleus. Since the nucleons are non-relativistic objects, the averaging of $ r_{i}^{2} $ over the vector and scalar densities leads to identical results.
For the mean value one can take $\left \langle r_ {i}^{2} \right \rangle = 3/5 R_{A}^2$, where $ R_{A} = 1.2 ~ A^{1/3} ~ \textrm {fm} $.
The difference is that the vector charge is normalized to unity, while the scalar charge (\ref{v2}) is normalized to $ \left \langle 1_{s} \right \rangle $. In the calculation of the scalar charge radius associated with the structure of the nucleus, by integrating over the quark variables $ \left \langle 1_{s} \right \rangle $ is factorized. The vector and scalar form factors at zero momentum transfer, when they are calculated using the non-relativistic wave functions of nucleons, are normalized to $ Z $ and $ Z \left \langle 1_{s} \right \rangle $, respectively.

Equation (\ref{GEQ114}) also holds for nuclear $J = I = 1/2 $ isotopic doublets. In general  \cite{PEPAN2015}
\begin{eqnarray}
\frac{g_{v+s}}{m_{\mu }} &=&-\frac{m_{I_{3}=+1/2} - m_{I_{3}=-1/2}}{6}\left[
\left( \left\langle r^{2}\right\rangle _{I_{3}=+1/2}-\left\langle
r^{2}\right\rangle _{I_{3}=-1/2}\right) \right.   \nonumber \\
&&~~~~~~~~~~~~~~~~~~~~~~~~~~~~~~~-\left. \left(
\left\langle r_{s}^{2}\right\rangle _{I_{3}=+1/2}-\left\langle
r_{s}^{2}\right\rangle _{I_{3}=-1/2}\right) /\left\langle 1_{s}\right\rangle
\right]. \label{GEQ116}
\end{eqnarray}
Using relations (\ref{rv}) and (\ref{rs}), we find that the expression in brackets in Eqs.~(\ref{GEQ114}) and (\ref{GEQ116}) coincide.
To the zero-order expansion in $ (v/c)^2 $, where $ v $ is velocity of nucleons, the form factor $g_{v + s}$ is determined by the charge radii of nucleons and the mass difference of isotopes.

For the system $ (^{3} {\rm He}, ^{3} {\rm H}) $ Eq.~(\ref{GEQ116}) gives $ g_{v + s}^{{\rm th}} = 4.3 \times 10^{- 5} $,
that does not contradict to the result of $ g_{v + s}^{{\rm expt}} = {\rm -0.005} \pm {\rm 0.040} $, obtained from the analysis of muon capture in $^{3} {\rm He } $
\cite {Gazit2008}. In the case of $ (^{3} {\rm He}, ^{3} {\rm H}) $, testing the theoretical predictions requires an increase in the measurement accuracy by three orders of magnitude.

\vspace{15pt}

List of the $ J = I = 1/2 $ nuclei and estimates of the $ g_{v + s}^{{\rm th}} $, based on the equation (\ref {GEQ116}), are shown in Table 1.
For some isotopes, the longitudinal form factor is the order of magnitude greater than for $ (^{3} {\rm He}, ^{3} {\rm H}) $, however, those isotopes have short lifetimes.

Among the isospin doublets listed in Table 1 are states formed by a nucleon or a nucleon hole on top of the filled shells of nucleons ($ A = 0, 4, 12, 16, \dots $).
Isospin doublets $ A = 5, 11, 17, 27$ have spins different from $ 1/2$, so they are not discussed here.

\vspace{15pt}

The generalized Ward-Takahashi identity for the weak VC, that takes into account isotopic symmetry breaking, was used for the evaluation of longitudinal vector form factors in the weak decays of the neutron and eight nuclei with
$ I=J = 1/2 $.
The result is proportional to the mass difference between the final and initial states and, for small momentum transfers, the difference between Lorentz-vector and Lorentz-scalar charge radii of the members of isotopic doublet. To measure the form factor of the longitudinal component of the weak VC in the muon capture by atoms
increasing the accuracy of measurements is required by two-three orders of magnitude.

\section{ACKNOWLEDGMENTS}
This work was supported by Grant no. HLP-2015-18 of Heisenberg-Landau Program and
Grant no. 3172.2012.2 for Leading Scientific Schools of Russia.


\nocite{*}
\bibliographystyle{aipnum-cp}%

\end{document}